# Tuning the metamagnetic transition in the (Co, Fe)MnP system for magnetocaloric purposes

F. Guillou[a)] and E. Brück

Fundamental Aspects of Materials and Energy (FAME), Faculty of Applied Sciences, Delft University of Technology, Mekelweg 15, 2629 JB Delft, The Netherlands

[a)]Author to whom correspondence should be addressed. Electronic mail: F.guillou@tudelft.nl

The inverse magnetocaloric effect taking place at the antiferro-to-ferromagnetic transition of (Co,Fe)MnP phosphides has been characterised by magnetic and direct $\Delta T_{ad}$ measurements. In $Co_{0.53}Fe_{0.47}MnP$, entropy change of 1.5 $Jkg^{-1}K^{-1}$ and adiabatic temperature change of 0.6 K are found at room temperature for an intermediate field change ($\Delta B$= 1 T). Several methods were used to control the metamagnetic transition properties, in each case a peculiar splitting of the antiferro-to-ferromagnetic transition is observed.

## I. INTRODUCTION

Magnetic refrigeration is a promising cooling technique which might advantageously replace the traditional gas compression technology, since it has an higher efficiency and does not use refrigerant gases which are greenhouse gases[1]. The search for new materials presenting a significant magnetocaloric effect (MCE) is a key challenge for the development of this technology. In order to reach large MCE -either quantified as an entropy change ($\Delta S$) or as a temperature change ($\Delta T_{ad}$)- the use of the latent heat exhibited by first order transition (FOT) is of primary interest. Around room temperature, most of the studies generally focus on materials presenting a first order ferro-to-paramagnetic transition (FM-PM), as for instance in the "giant" MCE compounds: $Gd_5(Ge,Si)_4$,[2] $Mn(As,Sb)$,[3] $La(Fe,Si)_{13}$[4] and its hydrides[5] and $Mn_{2-x}Fe_x(P,As,Ge,Si)$[6-9]. Far less attention is generally paid to materials having their MCE around an antiferro-to-ferromagnetic transition (AF-FM), the so-called "inverse" MCE ; though this strategy can also lead to a significant effect since these transitions are usually of first order type. Around room temperature, only a very few inverse MCE materials have been identified and, unfortunately, most of them are far from being promising for applications: (i) $Fe_{0.49}Rh_{0.51}$ is the MCE materials with the highest $\Delta T_{ad}$ measured so far, however the cost of Rh prohibits any large-scale use[10] ; (ii) In tetragonal Mn alloys, $Mn_{2-x}(Cr,V)_xSb$, the toxicity of antimony could be a difficulty[11] ; (iii) The Heusler alloys represent a rich family of MCE materials which can exhibit either a normal or an inverse MCE. Of particular interest are those of formulation Ni-(Mn,Co)-X (X=In,Ga, Sn), for which the transition separates a ferromagnetic austenite from a "less" magnetic martensite[12-14]. However, due to the large hysteresis generally observed in these compounds, it is still unclear if their MCE can really be used in a refrigeration cycle ; (iv) Finally, the last family is the silicates derived from CoMnSi.[15-18] Among the inverse MCE materials, the CoMnSi family appears to be one of the most promising. However, some features limit their MCE performances. In the parent CoMnSi composition, no significant MCE is found for a field variation of 1 T or 2 T, *i.e.* precisely in the intermediate fields range which is the most relevant for applications. In order to restore a MCE, substitutions can be used, but that shifts the Neel temperature far below room temperature.[16]

Following this unusual "inverse MCE" approach, attention has been paid during this study to others MnM'X alloys (with M'=3d and X=non-metal elements) that could present an inverse MCE. We especially focused on the phosphides $Co_{1-x}Fe_xMnP$ that have often been considered as archetypical examples of the AF-FM behavior in orthorhombic manganese alloys ($Co_2P$-type) and which are anticipated to have both a normal MCE at $T_C$ and an inverse at $T_N$.[19-21] In these materials, the occurrence of the AF- FM transition can be expected above half-substitution (x> 1/2), since on one hand the FeMnP parent composition was reported to be antiferromagnetic up to $T_N$ ~340 K,[22] while on the other hand, CoMnP is a ferromagnet with a relatively high $T_C$ (~600 K).[23] This series $Co_{1-x}Fe_xMnP$ was experimentally investigated for the first time in the 70's [24](a copy of the phase diagram can be found in Ref. 19). For x values between 0.5 <x <0.8, a sequence of AF-FM and then FM-PM transitions was found to take place upon warming.

The aim of the present study is to carry out an experimental investigation of the inverse MCE that can be expected around the AF-FM transition. This approach encouraged us to revisit the magnetic phase diagram of the $Co_{0.53}Fe_{0.47}MnP$ composition and to investigate new approaches to adjust the properties of the metamagnetic transition.

## II. EXPERIMENTAL DETAILS

Polycrystalline $Co_{1-x}Fe_xMnP$, $(Co_{1-x}Fe_x)_{1-y}Mn_yP$, $Co_{1-x}Fe_xMnP_{1-y}Si_y$, $Co_{1-x}Fe_xMnP_{1-y}Ge_y$ samples were prepared by using solid state reaction. Stoichiometric quantities of high purity precursors either in forms of powders (Co, Fe, Si, P) or chips (Ge, Mn) were first ball-milled during 8 hours. Then, the resulting powders were: pressed into pellets, sealed in quartz ampoules with 200 mbar Ar, sintered at 1273 K during 6 h, annealed at 1123 K during 72 h and finally cooled to room temperature with a slow rate of 0.2 Kmin$^{-1}$. The x-ray diffraction patterns were measured at room temperature with a PANalytical X-pert Pro diffractometer. The XRD patterns were refined by using the Fullprof software. SEM and EDX measurements were carried out in a Jeol JSM7500F microscope equipped with a Noran system. Magnetic measurements were carried out in a Quantum Design magnetometer equipped with a SQUID (MPMS 5S). Isothermal magnetization curves $M_T(B)$ were measured according to the following protocol. The sample was first zero field cooled to the measurement temperature and then two types of curves were consecutively recorded: (i) B is increased from 0 to $B_{max}$ and (ii) B is decreased to 0. A two steps process was also used to measure isofield magnetization curves $M_B(T)$. The magnetic field is applied at the highest temperature, the first measurement is performed upon cooling (FC mode) and then upon warming (FCW). These $M_B(T)$ measurements were performed in sweep mode at a temperature rate of 1.5 Kmin$^{-1}$. For the $\Delta S$ derivation, the field increment between each $M_B(T)$ curves is from 0.1 at low fields to 0.3 T in high fields, *i.e.* similar to the common field increment used for MCE calculation on the basis of $M_T(B)$ curves. The direct $\Delta T_{ad}$ measurements were performed in a home-made device. This measurement consists in inserting/removing the samples from a field generated by a permanent magnet, while measuring the temperature versus time signals by means of a thermocouple glued on the sample. The sample mass (2 g) being large compare to the thermocouple size, no thermal mass correction is applied. Quasi-adiabatic conditions are ensured during the $\Delta T_{ad}$ measurement since: (i) The samples are coated by an insulation layer, (ii) The magnetic field changes are performed with a high rate (1.1 T/s).

**III. RESULTS AND DISCUSSION**

**A. Transition lines in the $Co_{0.53}Fe_{0.47}MnP$ archetypical example**

In the $Co_{1-x}Fe_xMnP$ system, the x = 0.475 is an already known composition. The transition temperatures were initially found at $T_N \approx 295$ K and $T_C \approx 333$ K [19] (more recently, a lower AF-FM transition temperature has been reported $T_N \sim 195$ K in B = 0)[20]. In this study, a slightly different nominal composition with x= 0.47 will be investigated. The Figure 1a) displays a set of $M_B(T)$ curves for our x = 0.47 sample. In low magnetic field, B= 0.1 T, a raw estimate of the transition temperatures can be obtained by simply considering the middle point of the magnetization jumps. The transitions are located at $T_N \approx 305$ K and $T_C \approx 344$ K, *i.e.* shifted by 10 K in comparison to the values for the $Co_{1-x}Fe_xMnP$ x=0.475 of the litterature.[19,24] This discrepancy could be partially ascribed to the experimental methods and the criteria used to derived these transition temperatures, as well as to the slightly higher Cobalt content since a lower x value will lead to a stabilisation of the ferromagnetic phase, *i.e.* a decrease of $T_N$ and an increase of $T_C$. As a matter of fact, we can consider that this new sample display magnetic properties well in line with the previous report.[19] For sake of clarity and due to the closeness with the x=0.475 of the literature, our x=0.47 sample will be considered directly comparable hereafter. In B= 0.5 T, one observes that the AF-FM transition is sharp, the magnetization jump is of about 50 $Am^2kg^{-1}$ and takes place over a limited temperature range of 15 K. When the magnetic field is increased in the range 0.1 T ≤ B ≤ 0.7 T, as can be expected for a AF-FM transition, the $T_N$ is simply shifted toward lower temperatures. However, when one further increases the field, an original behaviour appears. For fields higher than 1 T, a splitting of the $T_N$ can be perceived. This evolution being relatively vague, a dM/dT(T) plot is used, Fig.1(b). Between 0.8 and 1 T, a splitting into two parts of the dM/dT peak can be clearly noticed, the critical field of this splitting is $B_{cri}$= 0.9 +/- 0.1 T. For B >$B_{cri}$, one observes that the low temperature transition, thereafter called $T_{N1}$, evolves much faster with the field than the "high temperature" transition $T_{N2}$. It deserves to mention that the splitting of the $T_N$ observed during isofield measurements $M_B(T)$ is consistent with the isothermal magnetization measurements $M_T(B)$. On Fig.1(c), three $M_T(B)$ curves corresponding to three specific temperature ranges of the phase diagram are shown : (i) At T= 280 K, one observes only one magnetization jump for B= 0.6 T which is attributed to $T_N$ and is in line with the single dM/dT peak found in Fig.1(b) for this field value ; (ii) At low temperature (220 K), only the transition with the highest dT/dB is crossed, *i.e.* $T_{N1}$, here at $B_{TN1}$ = 1.8 T ; (iii) And in between, at T= 250 K, the two transitions occur consecutively, a first magnetization jump (at $B(T_{N1})$= 1.2 T) clearly exhibits $T_{N1}$, while $T_{N2}$ appears as a broad curvature change (centered at about B= 3.1 T).

On the basis of magnetic data only, it is always difficult to attribute such a splitting of a transition to an intrinsic property or to the presence of secondary phases/inhomogeneities, especially in these Mn-alloys for which the vapour pressure of Mn and the volatility of phosphorous can easily lead to an off-stoichiometry. In order to clarify this point, EDX analyses were carried out on different spots of a bulk piece having a nominal composition $Co_{0.53}Fe_{0.47}MnP$. Fig.2(a) shows a typical EDX mapping for one of these spots, the Mn, Fe, Co and P elements appear to be well spread in the material (similar observations were made on the other spots). The atomic proportions are 33.4 (3) % of Mn, 33.1 (2) % of P, 16.0 (3) % of Fe and 17.4 (3) % of Co, which is in agreement with the nominal composition. The x-ray diffraction pattern of this x=0.47 material is shown in Fig.2(b) and all the peaks can be indexed in the P*nma* space group. The lattice parameters a= 5.9429(3) Å, b= 3.5519(2) Å and c= 6.7403(4) Å are close to the results for x=0.475,[19] the cell volume of this x=0.47 material

V= 142.278 Å$^3$ is slightly lower than in the x=0.475 of the literature due to the higher Co content. EDX and XRD results thus strongly suggest that this splitting of the AF-FM transition is an intrinsic feature of the $Co_{1-x}Fe_xMnP$ materials.

By considering the transition temperatures ($T_N$ and $T_C$ for B≤ 0.8 T) ($T_{N1}$, $T_{N2}$ and $T_C$ for B ≥ 1T) as the maxima (or minima for $T_C$) on the dM/dT curves from Fig.1(b), it is possible to construct the magnetic phase diagram displayed on Fig.3. A set of $M_B(T)$ curves were also recorded upon cooling –FC- (not shown). A limited thermal hysteresis is only noticeable around the $T_{N2}$ transition (with $\Delta T_{hyst}$= 1 K in B= 1 T to $\Delta T_{hyst}$= 3 K in B= 5 T). Accordingly, to draw the B-T plot, only one transition point has been considered for each transition. In the magnetic phase diagram, the splitting of $T_N$ leads to the opening of a pocket which grows with the field. It deserves to mention that there is one order of magnitude between field evolution of $T_{N1}$ and $T_{N2}$, such a large difference seems rather incompatible with minor off-stoechiometries, impurities or inhomogeneities that would not be detected by XRD or EDX experiments.

In comparison to the AF- FM transition line already reported[20], two major discrepancies appear: (i) in the present study, no anomaly was found around 195 K in B=0 ; (ii) the transformation from the antiferromagnetic ground state to ferromagnetism occurs in two steps for B ≥ 1 T, *i.e.* one observes the existence of an additional transition. It is not unreasonable to consider that this splitting of the AF-FM above 0.9 T was not observed during this previous study due to an insufficient resolution. Actually, it must be emphasized that $T_{N2}$ is found only over a restricted temperature range of the phase diagram (over 20 K), this range being much smaller than the 50 K temperature increments used during the previous study.

About the origin of $T_{N2}$, the problem is that even though magnetic measurements are useful to reveal transitions, they do not provide direct information about their natures. Let us however propose a scenario based on qualitative observations. The (Co,Fe)MnP phosphides share a lot of similarities with the CoMnSi based materials. In these silicates, a similar splitting of the $T_N$ has already been observed in intermediate magnetic fields, and attributed to a two steps transition from the AF ground-state (helical) to a ferromagnetism through an intermediate fan structure.[25] Assuming that $Co_{0.53}Fe_{0.47}MnP$ has the same AF ground state of helical type as his FeMnP parent[26], a similar scenario can be considered: for B < $B_{cri}$ upon warming, only $T_N$ (Helical to Ferromagnetic) and $T_C$ (Ferro to Paramagnetic) transitions are crossed ; while for B> $B_{cri}$ the sequence of transitions upon warming corresponds to $T_{N1}$ (Helical to fan), then $T_{N2}$ (fan to FM) and $T_C$ (FM to PM).

To support this scenario, one can notice that $Co_{0.53}Fe_{0.47}MnP$ shares another similarity with the silicates derived from CoMnSi, the appearance of a finite latent heat (L) that grows with increasing the field.[16,17] In order to highlight this feature, the original interpretation of the Clausius-Clapeyron equation is used, which is the estimate of the entropy change of a FOT along its transition line: $\Delta S_{tr} = L/T_{tr} = \Delta M (dB/dT)_{tr}$. This Clausius-Clapeyron analysis is usually made delicate by the choices in the criteria used to collect $\Delta M$ values. If around a first order FM-PM transition, $\Delta M$ can be easily estimated by using two parallels lines on each side of $T_C$, in an AF-FM case, the opposite $M_B(T)$ slopes before/after the transition prohibit this method. Here, it has been considered that the AF magnetization is given by a linear extrapolation towards higher temperatures. For the upper boundary of $\Delta M$, the proximity between $T_N$ and $T_C$ induces a certain rounding. To take it into account, the curve M(T, B= 5 T) scaled by a factor is used to estimate the FM magnetization at T= $T_{N2}$ for B< 5 T. To estimate the magnetization values at the $T_{N1}/T_{N2}$ boundary (when B > $B_{cri}$), the simplest case -a horizontal line- is chosen which corresponds to attribute at each transitions a part of the whole magnetization jump. The $\Delta S_{tr}$ derived along each AF-FM transition line ($T_N$, $T_{N1}$ and $T_{N2}$) are displayed in Fig.4(a). The values obtained for $T_N$ ($\Delta S_{tr}$ ~ 0.4 $Jkg^{-1}K^{-1}$) and $T_{N1}$ ($\Delta S_{tr}$ ~ 0.3 $J kg^{-1} K^{-1}$) are low and do not show any pronounced temperature evolution, while the entropy change of $T_{N2}$ corresponds to a finite latent heat that rapidly grows with increasing the field (decreasing the temperature). It should be noticed that the $\Delta S_{tr}$(B= 0.1 T) = 0.3 $Jkg^{-1}K^{-1}$ obtained from the Clausius-Clapeyron method in low fields is in good agreement with the $\Delta S_{tr}$ = 0.26 $Jkg^{-1}K^{-1}$ obtained by DSC measurements in B= 0.

A first direct consequence of the $T_N$ splitting and $\Delta S_{tr}$ evolution is that a significant MCE will be obtained only when the field is large enough to involve the latent heat of $T_{N2}$ into the MCE, *i.e.* only when the magnetic field change $\Delta B$ is larger than $B_{cri}$. At first glance to optimize these materials for MCE applications, it is required to reach the highest $T_{N2}$ latent heat values for one given field change, that is to say, one has to shift the $B_{cri}$ value toward lower fields. However, the situation is made more complex by the rapid increase of the width of the AF-FM transition (the whole transition, from the ground-state to FM) in high fields. This last outcome of the $T_N$ splitting is shown on Fig.4(b), where the increase of the width of the AF-FM transition *vs* B shows a strong acceleration above $B_{cri}$.

### B. Tuning of the AF-FM transition

For MCE purposes, it is of interest to find parameters to manipulate the metamagnetic behaviour. Not only for the obvious reason of controlling the AF-FM transition temperature, but also because, as noted before, the $B_{cri}$ value will have a significant influence on the MCE performances. A first method to tune the metamagnetic transition is to explore the $Co_{1-x}Fe_xMnP$ phase diagram in the range x< 0.47. Here, in order to have a more general approach two additional methods were also investigated: variations in the Mn content $(Co_{1-x}Fe_x)_{1-y}Mn_yP$ and partial substitutions of phosphorus by elements of the IVA group. Though a large variety of composition is conceivable, only a few representative examples corresponding to modulations of the x= 0.47 and x= 0.3 $Co_{1-x}Fe_xMnP$ compositions will be presented. The Neel temperatures (derived from $M_B(T)$ curves recorded in B= 0.1 T) are reported in Fig.5 as a function of the shortest distance between the pyramidal sites (assumed to be occupied by Mn atoms). In the present study carried out at room temperature, *i.e.* in the ferromagnetic state, the shortest Mn-Mn distance corresponds to the distance between Mn atoms in the y= 1/4 and y= 3/4 planes. This $d_{Mn-Mn}$ criterion was chosen, since several authors have already emphasized the importance of distances between metallic elements to account for the magnetic properties of MnM'X compounds.[22,23,27,28] Moreover, it has recently been suggested that this parameter plays the dominant role in determining the magnetic properties of these compounds.[21] First, our $d_{Mn-Mn}$ values are well consistent with known boundaries: the $d_{Mn-Mn}$ are higher than in CoMnP (2.83 Å) which has a ferromagnetic ground state and lower than in FeMnP (3.05 Å) which is an antiferromagnet.[23] Second, these experimental results are well consistent with theoretical predictions; the AF-FM transition is only obtained over a limited range of $d_{Mn-Mn}$ values (around 2.9 Å).[21] Third, though an universal $T_N$ *vs* $d_{Mn-Mn}$ curve is not obtained, a general tendency, an increase of $T_N$ with $d_{Mn-Mn}$ is found. More precisely for each type of substitution, one observes that: (i) Starting from $Co_{0.7}Fe_{0.3}MnP$, the $T_N$ is increased by replacing P by Si or Ge. The same substitutions in $Co_{0.53}Fe_{0.47}MnP$ also lead to a shift of $T_N$ toward higher temperatures and more significantly induce a weakening of the ferromagnetic phase (lower magnetization values above $T_N$ than in the parent compound). (ii) The $T_N$ is driven towards higher temperatures either by increasing the Fe/Co ratio or the Mn/(Fe,Co) ratio. Thus, the AF-FM transition temperature is easily controlled by structural parameters. In particular, in all the investigated substitutions, it is found that the insertion of bigger elements increases both $d_{Mn-Mn}$ and $T_N$. We suggest that this $d_{Mn-Mn}$ criterion is not only a useful parameter to predict the existence of the metamagnetic transition, but it could also provide rules to adjust the AF-FM transition temperature within this family of materials. Figure 6(a) shows the AF-FM transition lines for some selected compositions that can be compared with our archetypical x= 0.47 example displayed in Fig.3. In all these cases, a splitting of $T_N$ is observed. For $Co_{0.55}Fe_{0.45}MnP$, $B_{cri}$ is 0.9 +/- 0.1 T and as expected very similar to x= 0.47. For $Co_{0.7}Fe_{0.3}MnP_{0.95}Ge_{0.05}$ and $Co_{0.7}Fe_{0.3}MnP$, the $B_{cri}$ are 0.8 T and 0.5 T, respectively. The "splitting" field $B_{cri}$ turns out to be hardly modified by substitutions, whereas the associated critical temperature decreases as rapidly as the Neel temperature ($T_N$). No precise rules were found about the $B_{cri}$ evolution. In this set of samples, when $T_N$ is decreased, $B_{cri}$ also decreases, a qualitative evolution which is also observed in the CoMnSi based materials.[25] An additional indication of the $B_{cri}$ decrease when $T_N$ is decreased is the appearance of hysteresis in intermediate field which results -at least partially- from a more pronounced first order character of $T_{N2}$ in B= 1T. As can be seen in Fig.6(b), contrary to x= 0.47 or 0.45, a finite thermal hysteresis (10 K) is observed for $Co_{0.7}Fe_{0.3}MnP$.

**C. Magnetocaloric properties**

The Maxwell equation $\Delta S(T; \Delta B) = \partial [\int_0^B M\, dB']/\partial T$ has been used to derive the isothermal entropy changes from the isofield magnetization curves recorded upon warming. The results for $Co_{0.53}Fe_{0.47}MnP$ are plotted Fig.7(a) in comparison with the compositions from the Fig.6(a). For a field change of $\Delta B = 1$ T, one can distinguish two $\Delta S(T)$ shapes. When the $B_{cri}$ of a sample is about the applied field change (x= 0.47 and x= 0.45 cases), one observes a small kink on the low temperature side of the main $\Delta S$ peak which corresponds to $T_{N1}$. While when the $B_{cri}$ is significantly lower than 1 T (x= 0.3 and $Co_{0.7}Fe_{0.3}MnP_{0.95}Ge_{0.05}$), the $T_{N1}$ leads to an anomalous enlargement of the $\Delta S$ peak on its low temperature wing. At a given field change (in our study $\Delta B > B_{cri}$), compounds with a lower $B_{cri}$ exhibit lower $\Delta S$ values, but over a larger temperature range. This behaviour is related to the fast increase of the width of the AF- FM transition after crossing the splitting point.

A "normal" MCE is expected around the Curie temperature. In $Co_{0.53}Fe_{0.47}MnP$, this MCE has a maximum of -1.5 $Jkg^{-1}K^{-1}$ at 356 K for $\Delta B$= 2 T, which is lower than the +2.0 $Jkg^{-1}K^{-1}$ found around AF-FM transition. The inverse MCE of this MnM'X phosphides are modest, the absolute $\Delta S$ values are of the same order of magnitude as what is commonly found in second order materials around room temperature, for instance in manganites.[1] For an intermediate field change ($\Delta B$= 2 T), these $\Delta S^{max}$ are significantly higher than in the original CoMnSi compound ($\Delta S^{max}$~ 0.2 $Jkg^{-1}K^{-1}$)[15] and similar to the optimized compositions $CoMnSi_{1-x}Ge_x$ ($\Delta S^{max}$~ 1.1 $Jkg^{-1}K^{-1}$)[15] or $CoMn_{0.95}Fe_{0.05}Si$ ($\Delta S^{max}$~ 3.0 $Jkg^{-1}K^{-1}$)[16].

The second MCE quantity, the adiabatic temperature change, was measured by a direct method consisting in monitoring the sample temperature while field oscillations are applied, see inset Fig.7(b). The external temperature being swept (0.5 K/min), a $\Delta T_{ad}$ versus T curve can be built, Fig.7(b). It should be noted that since the sample is continuously enduring magnetization/demagnetization cycles, these $\Delta T_{ad}$ values correspond to a reversible MCE. In $Co_{0.53}Fe_{0.47}MnP$, a maximal $\Delta T_{ad}$ of 0.65 K is obtained at 280 K for $\Delta B= 1.1T$. This value is significantly lower than the 2.5 K/T of the reference MCE material which is Gd and lower than most of the "Giant" MCE materials.[1] However, the hysteresis being only noticeable in high fields, the $\Delta T_{ad}$ in intermediate field of $Co_{0.53}Fe_{0.47}MnP$ is fully reversible, inset Fig.7(b), and is found to be very close to the reversible $\Delta T_{ad}$ of some "promising" inverse MCE materials. For instance in comparison to Heusler alloys, the $\Delta T_{ad}$ of $Co_{0.53}Fe_{0.47}MnP$ is higher than the $\Delta T_{ad}= 0.7$ K for $\Delta B= 3$ T found in $Ni_{51.3}Mn_{32.9}In_{15.8}$ [29] and similar to the reversible $\Delta T_{ad}= 1.3$ K for $\Delta B= 1.9$ T observed in $Ni_{45.2}Mn_{36.7}In_{13}Co_{5.1}$.[30] Finally, it should be noted that the $\Delta T_{ad}$ peak covers a large temperature range, with a width at half maximum of 40 K, which could be useful for applications since this leads to a significant cooling capacity.[1]

Regarding the $\Delta T_{ad}$ value itself, it should be noted that the $\Delta T_{ad}$ is considerably lower than the shift of the transition due to the field $\Delta T_{tr}$ ~40 K for $\Delta B= 1.1$ T. This can be regarded as a further evidence of the complex role played by the splitting of the AF-FM and the negative effect of the transition width on the MCE.

**IV. CONCLUSION**

The inverse magnetocaloric effect associated to the AF-FM transition taking place in $Co_{1-x}Fe_xMnP$ phosphides has been studied. It appears that the splitting of the AF-FM transition in our bulk samples of $Co_{0.53}Fe_{0.47}MnP$ and related materials plays an important role on the MCE performances, since the splitting point corresponds to the onset of a first order transition $T_{N2}$ having a finite latent heat and to a rapid increase of the width of the AF-FM transition.

Starting from the prototypical composition $Co_{0.53}Fe_{0.47}MnP$, several parameters –Fe/Co ratio, (Fe,Co)/Mn ratio and substitutions of P by IVa elements- were used to control the metamagnetic transition. In all these cases, a splitting of the AF-FM transition in intermediate field was observed and the critical field of this splitting decreases with the lowering of $T_N$ by substitutions.

Finally, despite modest $\Delta S$ and $\Delta T_{ad}$ maximal values, $Co_{0.53}Fe_{0.47}MnP$ turns out to be an interesting MCE material, this is actually the only orthorhombic MnM'X alloy having a significant inverse MCE at room temperature in intermediate fields and the effect is fully reversible. Obviously, further improvements of the MCE performances in these materials will require more experimental/theoretical work about the origin of the splitting of the AF-FM transition.

ACKNOWLEDGEMENTS

This work was supported by the Dutch Foundation for Fundamental Research on Matter (FOM) and by BASF Future Business.

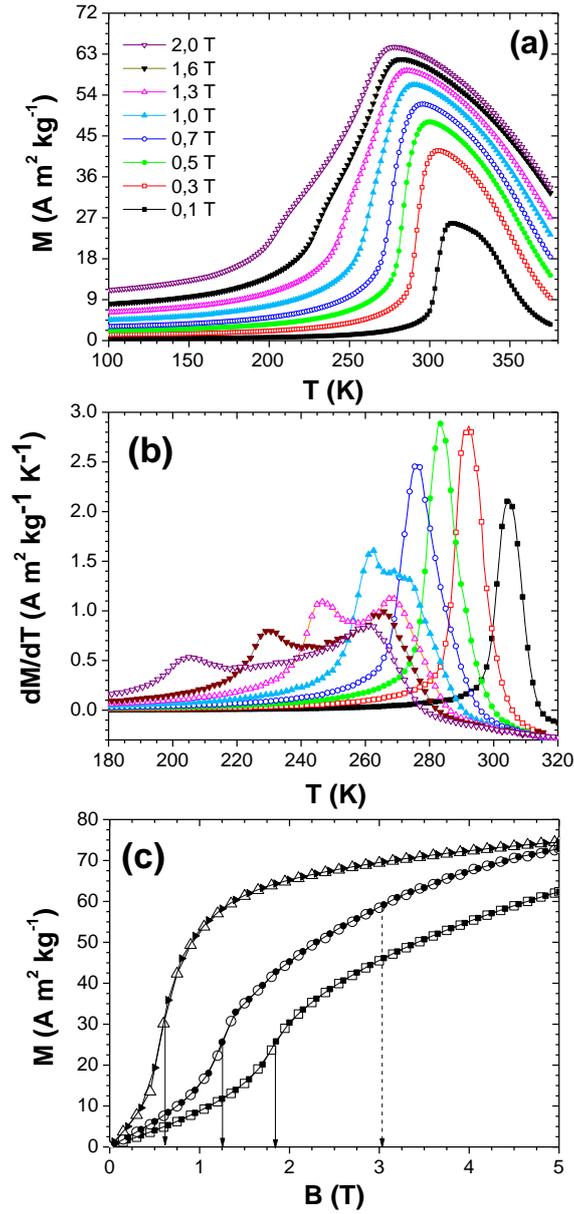

Fig.1. Magnetization properties of $Co_{0.53}Fe_{0.47}MnP$. (a) $M_B(T)$ curves in various magnetic fields recorded with a FCW protocol. (b) Derivative (dM/dT) of the curves from figure (a). (c) $M_T(B)$ curves recorded upon field increasing (open symbols) and field decreasing (closed symbols) at various temperatures: 280 K (triangles) ; 250 K (circles) ; 220 K (squares).

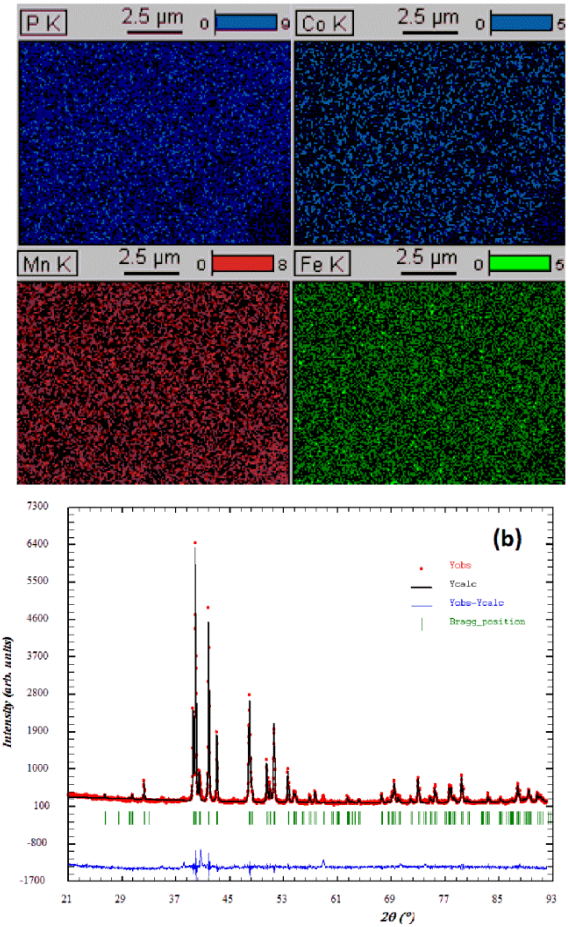

Fig.2. Top panel, EDX mapping of the different Mn, Fe, Co and P elements on a Co$_{0.53}$Fe$_{0.47}$MnP bulk piece. (b) X-ray diffraction pattern of Co$_{0.53}$Fe$_{0.47}$MnP.

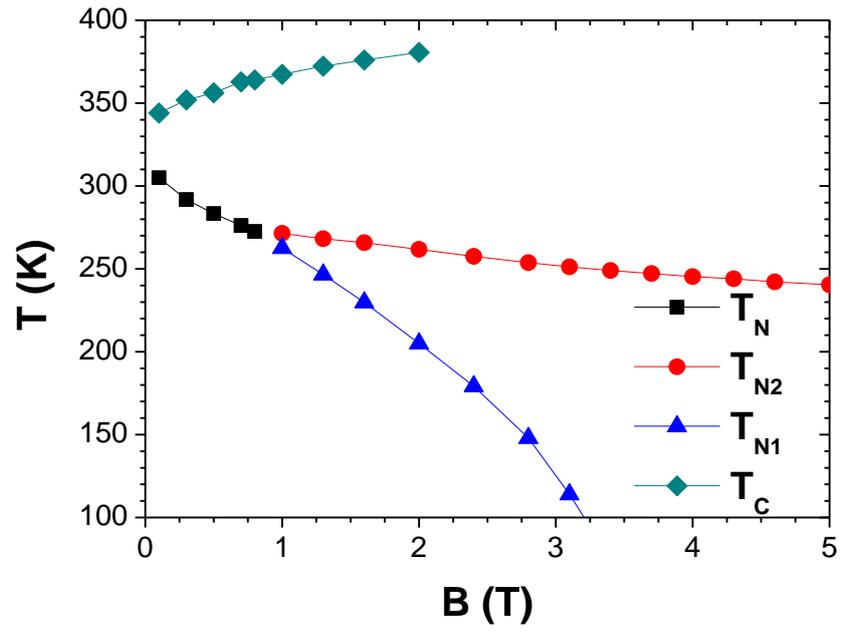

Fig.3. Magnetic field-temperature phase diagram of $Co_{0.53}Fe_{0.47}MnP$ derived from $M_B(T)$ measurements.

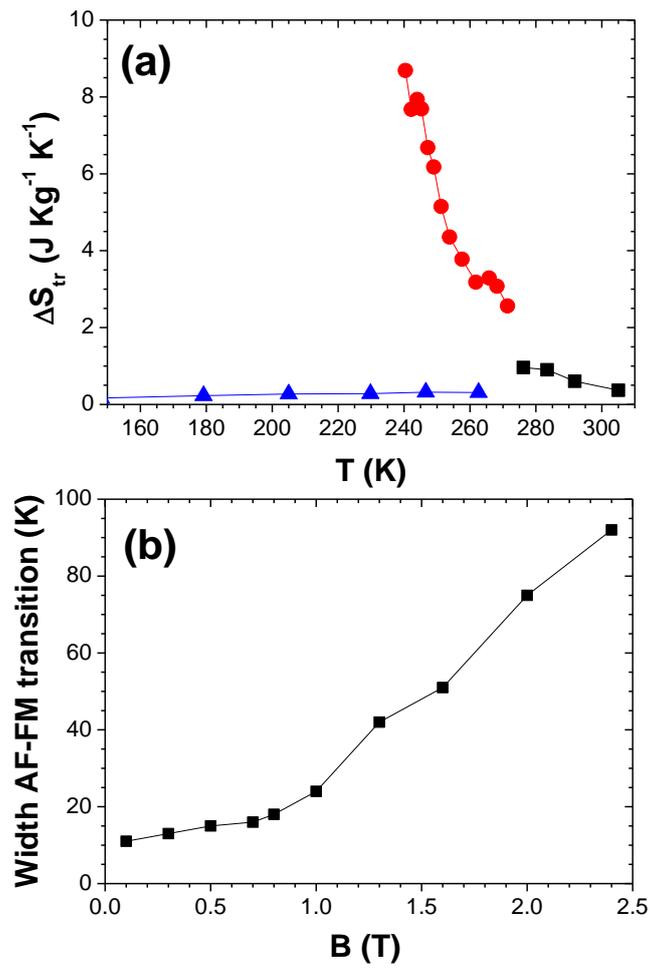

Fig.4. (a) Entropy changes $\Delta S_{tr}$ derived from the Clausius-Clapeyron method around the transitions in $Co_{0.53}Fe_{0.47}MnP$ : $T_N$ (squares) ; $T_{N1}$ (triangles) ; $T_{N2}$ (circles). (b) Width of the AF-FM transition as a function of the applied field.

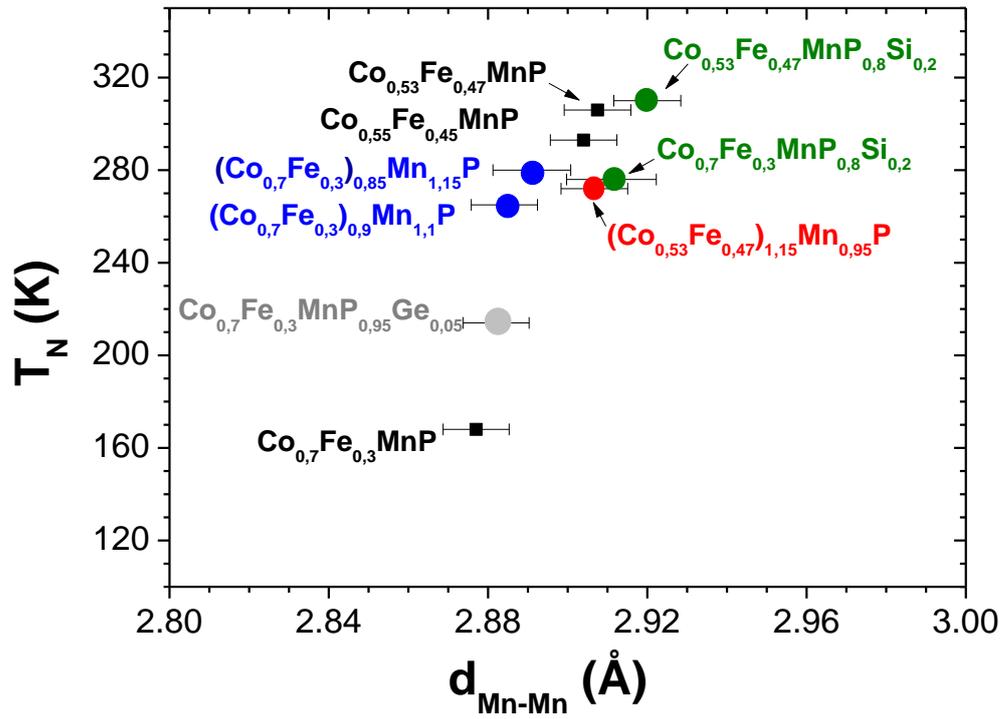

Fig.5. Neel temperatures (derived from $M_B(T)$ measurements in B= 0.1 T) of (Co,Fe)Mn(P,X) compounds as a function of the shortest Mn-Mn distance (obtained from XRD measurements carried out at 300 K).

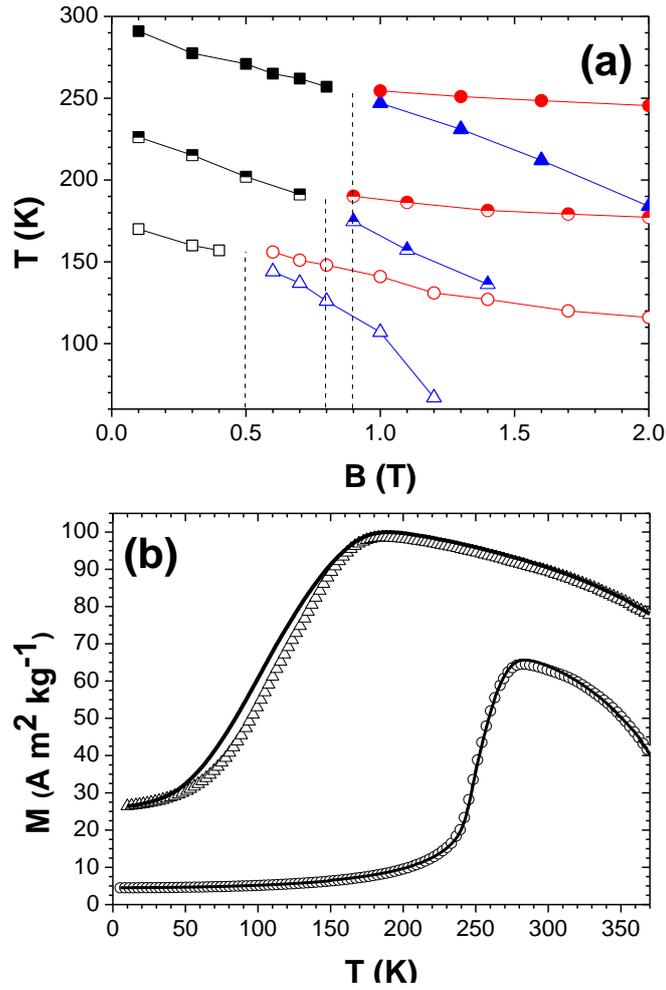

Fig.6. (a) Transition lines: $T_N$ (squares) ; $T_{N1}$ (triangles) ; $T_{N2}$ (circles) of $Co_{0.55}Fe_{0.45}MnP$ (filled symbols), $Co_{0.7}Fe_{0.3}MnP_{0.95}Ge_{0.05}$ (half-filled symbols) and $Co_{0.7}Fe_{0.3}MnP$ (open symbols). (b) Temperature dependence of the magnetization in a magnetic field of 1 T, upon cooling (line) and warming (open symbols) for $Co_{0.55}Fe_{0.45}MnP$ (circles) and $Co_{0.7}Fe_{0.3}MnP$ (triangles).

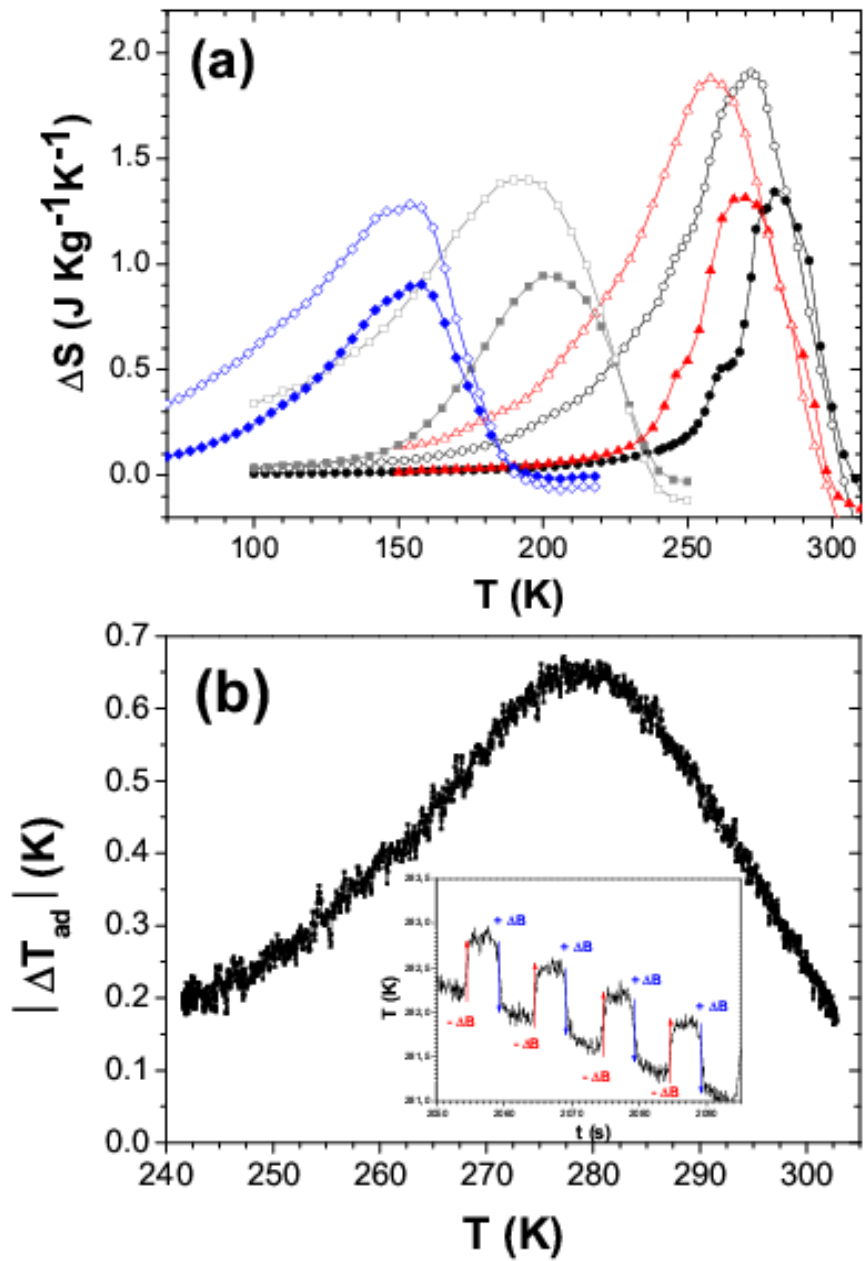

Fig.7. (a) Isothermal entropy changes for ΔB = 1 T (filled symbols) and ΔB = 2 T (open symbols) in: $Co_{0.53}Fe_{0.47}MnP$ (circles) ; $Co_{0.55}Fe_{0.45}MnP$ (triangles) ; $Co_{0.7}Fe_{0.3}MnP_{0.95}Ge_{0.05}$ (squares) ; $Co_{0.7}Fe_{0.3}MnP$ (diamonds). (b) Adiabatic temperature changes of $Co_{0.53}Fe_{0.47}MnP$ measured by a direct method for a field change of ΔB=1,1 T.